\documentclass[lettersize,journal]{IEEEtran}
\usepackage{amsmath,amsfonts}
\usepackage{algorithm}
\usepackage{array}
\usepackage[caption=false,font=normalsize,labelfont=sf,textfont=sf]{subfig}
\usepackage{textcomp}
\usepackage{stfloats}
\usepackage{url}
\usepackage{verbatim}
\usepackage{graphicx}

\usepackage{svg}
\usepackage{enumitem}
\usepackage{soul}

\usepackage{algpseudocode}
\usepackage[style=ieee,backend=biber]{biblatex}

\usepackage{eurosym}
\usepackage{amssymb} 
\usepackage[normalem]{ulem}
\usepackage{booktabs}

\usepackage{multirow}
\usepackage{makecell}
\usepackage{bm}
\usepackage[table]{xcolor}

\definecolor{citepurple}{HTML}{A4ACBB}
\definecolor{valgreen}{HTML}{8DCB20}
\definecolor{relgreen}{HTML}{ECF2FF}
\definecolor{customgray}{HTML}{E6E6E6}

\usepackage[colorlinks=true,
            linkcolor=black,
            urlcolor=citepurple,
            citecolor=citepurple]{hyperref}
\usepackage{cleveref}  
\begin{document}

\title{A Market-Rule-Informed Neural Network for Efficient Imbalance Electricity Price Forecasting}


\author{Runyao Yu$^{1,2}$, 
Julia Lin$^{2}$, 
Derek W. Bunn$^{3}$, 
Jochen Stiasny$^{1}$, 
Wentao Wang$^{4}$, 
Yujie Chen$^{5}$, 
Tara Esterl$^{2}$, \\
Peter Palensky$^{1}$,
Jochen L. Cremer$^{1,2}$\\ 
$^1$Delft University of Technology, $^2$Austrian Institute of Technology, 
$^3$London Business School,\\
$^4$University of Technology Sydney,
$^5$The Chinese University of Hongkong (Shenzhen)}

\maketitle

\begin{abstract}
Accurate and efficient imbalance electricity price forecasting is critical for industrial energy trading systems, especially as battery assets and automated bidding pipelines increasingly participate in balancing markets. However, real-time forecasting is complicated by nonlinear market-rule-based price formation, heterogeneous input signals, and incomplete data availability caused by communication delays, publication lags, and measurement outages. This paper proposes a market-rule-informed neural forecasting framework that embeds imbalance price formation rules into the latent space of an expressive neural network. The proposed framework preserves raw signal information while exploiting transparent market-rule priors. We further analyze operational robustness by removing price-component information and characterize how forecasting performance scales with input length and forecasting horizon. Experimental results show that the proposed model achieves competitive forecasting performance with substantially fewer trainable parameters and shorter training time than generic deep learning baselines. 
Experimental results show that the proposed model achieves competitive forecasting performance with substantially fewer trainable parameters and shorter training time than generic deep learning baselines, demonstrating that market-rule priors and expressive neural networks should be jointly used for accurate and \textcolor{black}{computationally sustainable} forecasting in industrial energy trading applications.
The implementation is publicly available at \textcolor{citepurple}{\url{https://runyao-yu.github.io/MRINN/}}.
\end{abstract}


\section{Introduction}

Electricity markets play a central role in modern power systems by coordinating generation scheduling, flexibility procurement, and real-time system balancing. In most liberalized systems, trading is organized sequentially across multiple layers: the day-ahead market clears the bulk of energy transactions one day before delivery, the intraday market allows participants to update positions closer to real time, and the balancing market resolves residual imbalances during system operation. 
Together, these markets form a hierarchical structure that links economic efficiency to physical system reliability.
\textcolor{black}{In particular, balancing-market operation involves two closely related mechanisms: participants may provide balancing services through accepted bids and offers, or hold imbalanced positions that are settled at the imbalance price~\cite{bunn2021statistical}.}

Imbalance prices are determined by predefined settlement rules that map observable system signals, such as system imbalance volumes and activated reserve energy, to an imbalance price. 
\textcolor{black}{However, these rules are not universal: different countries and bidding zones, such as Germany and Austria, adopt different imbalance settlement mechanisms. }
Although the pricing formulation is explicitly specified by market design and publicly available, forecasting imbalance prices remains challenging due to nonlinear and piecewise rule structures, scarcity adjustments, and rapidly changing operating conditions~\cite{iv2, iv9, o2024electricity, 8398478}.

From an industrial perspective, imbalance price forecasting is also a real-time information-processing problem: heterogeneous system, market, and reserve signals must be collected, synchronized, and transformed into reliable forecasts under strict temporal constraints. This makes the problem closely related to data availability, rule-based information processing, computational efficiency, and deployability in operational energy systems~\cite{runyaoEEM}.

Recent advances in deep learning have stimulated a growing body of work on non-linear electricity price forecasting~\cite{iv6, 8693845, iv10, 11456228, iv11,10418046,  runyaopscc}. Most existing approaches treat imbalance prices as generic time-series targets and rely on purely data-driven models (e.g., LSTM or Transformer). However, in contrast to many other domains, imbalance price formation is governed by transparent and explicitly stated market rules. Surprisingly, much of the current literature does not exploit this structural information, implicitly requiring models to relearn deterministic mappings that are already known. This gap motivates a re-examination of how market-rule knowledge should be incorporated into forecasting models.

Since the imbalance price $P_t$ at time $t$ is deterministically computed from raw features $F_t$ through a \emph{known sequence of pricing rules}, which will be detailed in Section~\ref{sec:prelim}, we write it at a high level as $P_t=g(F_t)$. 
Thus, forecasting $P_{t+15\text{min}}$ at time $t$ can be approached in two fundamentally different ways: (i) directly using the raw features $F_t$, or (ii) compressing them into the  imbalance price $P_t$ and using the compressed price.
However, the mapping from $F_t$ to $P_t$ is \textbf{non-invertible}: different $F_t$ can lead to the same $P_t$. 
\textcolor{black}{For example:
\begin{itemize}
    \item $F_t=[1,0,1,\ldots] 
    \xrightarrow{g(\cdot)} P_t=100 
    \xrightarrow{\text{model}} P_{t+15\text{min}}=200$;
    \item $F_t=[1,2,3,\ldots] 
    \xrightarrow{g(\cdot)} P_t=100 
    \xrightarrow{\text{model}} P_{t+15\text{min}}=500$.
\end{itemize}
If a forecasting model only uses $P_t$ as input, these two cases become indistinguishable: the model receives the same input value ($P_t=100$), but is expected to predict two different future prices. As a result, such a model is likely to just learn an averaged prediction. }
This paper therefore investigates whether forecasting should rely on raw features $F_t$ or rule-based price compression via $P_t=g(F_t)$. 

Moreover, when using raw features, a key question is how to effectively exploit prior knowledge encoded in the price formation rules. To this end, we design a \textbf{Market-Rule-Informed Neural Network (MRINN)} that takes the raw signals as input and embeds the price formation rules as a differentiable operator in the latent space, thereby injecting structural inductive bias while avoiding information loss due to premature compression. This rule-based injection also prevents the model from \textbf{relearning} price formation mechanisms solely from data with heavy parameterization, making MRINN parameter-efficient and suitable for resource-constrained deployment.

\begin{figure*}[!t]
    \centering
    \includegraphics[width=1\textwidth]{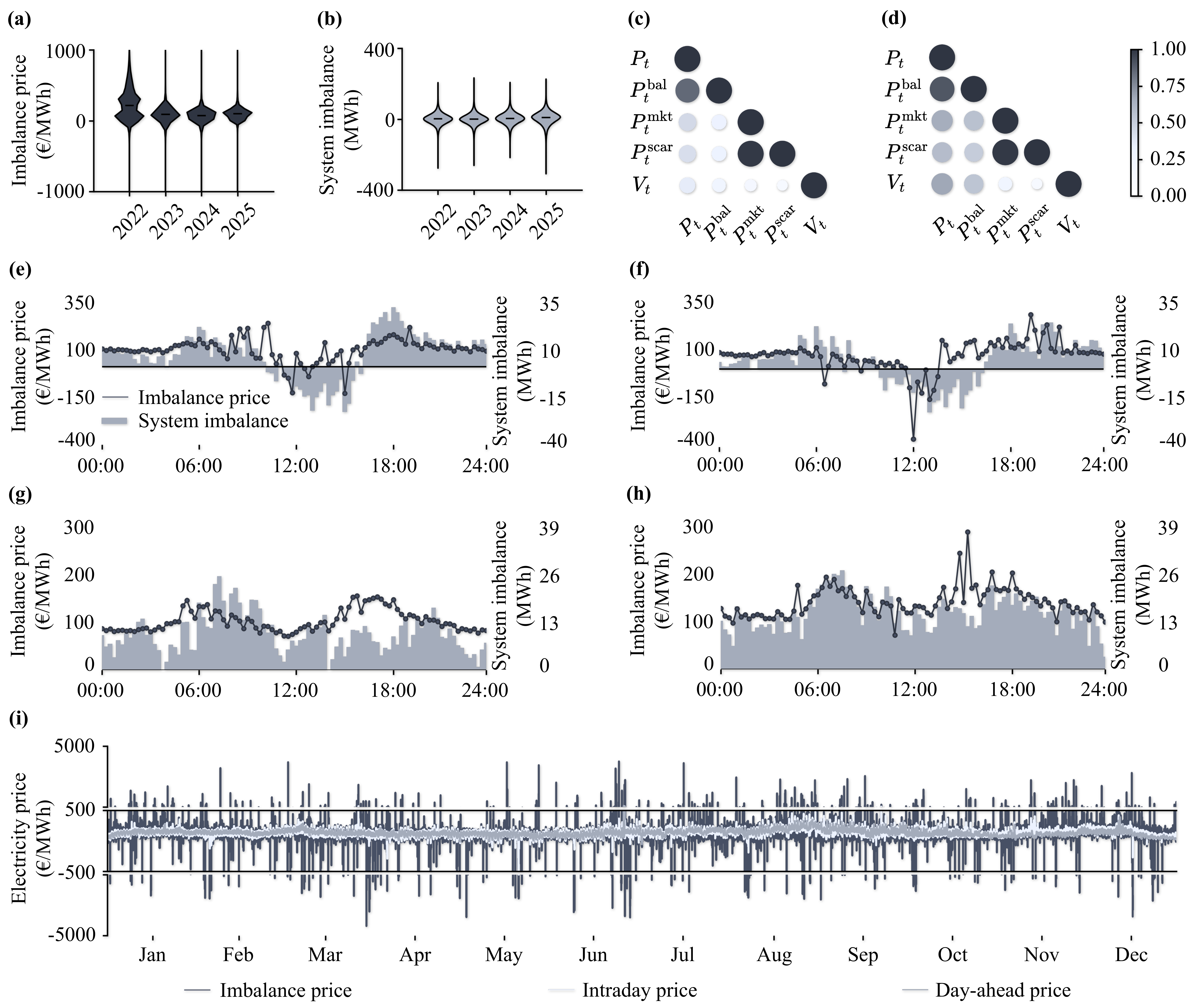}
    \caption{Imbalance data analysis. 
\textbf{(a)} Yearly distributions of imbalance prices from 2022 to 2025. 
\textbf{(b)} Yearly distributions of system imbalance from 2022 to 2025. 
\textbf{(c)} Pearson correlation magnitude among the imbalance price, the three price components, and system imbalance. 
\textbf{(d)} Spearman correlation magnitude among the same variables. 
\textbf{(e)--(h)} Average imbalance price and system imbalance across delivery times for spring, summer, autumn, and winter, respectively. 
\textbf{(i)} Comparison of day-ahead, intraday, and imbalance prices averaged across 2022--2025.
}
\label{fig:data_analysis}
\end{figure*}

Furthermore, in practical deployments, not all input signals are reliably available at time $t$ due to communication delays and occasional measurement outages. 
It therefore remains unclear how forecasting performance degrades when specific signals are consistently missing. Additionally, while prior work has investigated scaling laws in input length and forecasting horizon~\cite{deng2024seasonality, iv2}, it is not obvious whether such laws persist, or how they change, under models that explicitly incorporate price formation rules. In this paper, we systematically analyze the impact of different data availability and scaling behavior using MRINN.
Our main contributions are:
\begin{itemize}
    \item We propose a Market-Rule-Informed Neural Network (MRINN) that embeds imbalance pricing rules into the latent space with expressive data-driven learning while achieving resource-efficient probabilistic forecasting.

    \item We benchmark the proposed model against na\"ive, lagged-price-based, and raw-feature-based baselines to evaluate whether operational forecasting should rely on compressed lagged prices or raw industrial data streams.
    \item We analyze operational sensitivity by varying input availability and characterizing empirical scaling laws with respect to input length and forecasting horizon, providing actionable guidance
    for real-world deployment.
\end{itemize}

\textcolor{black}{
The remainder of this paper is organized as follows. Section~\ref{sec:data} introduces the dataset, providing generic insights. Section~\ref{sec:prelim} explains the imbalance pricing rules, which are subsequently injected into the proposed MRINN model in Section~\ref{MRINN}. Section~\ref{baselines} presents the baseline models. Section~\ref{operational} analyzes operational sensitivity under varying data-availability conditions. Section~\ref{experiment} reports the empirical results, and Section~\ref{conclusion} concludes the paper. }
\section{Data}
\label{sec:data}

The dataset is with the 15-min resolution and spans 4 years, from 2022-01-01 to 2026-01-01. The Fig.~\ref{fig:data_analysis} summarizes the empirical characteristics of the key features from the Austrian balancing market. 

\subsection{Feature Categories}
The input features can be grouped according to the imbalance pricing structure, which will be introduced in Section~\ref{sec:prelim}.

The first group contains balancing-energy features\textcolor{black}{, including automatic Frequency Restoration Reserve (aFRR) and manual Frequency Restoration Reserve (mFRR) activation volumes and corresponding activation prices. 
Although aFRR and mFRR are European balancing-service products, analogous real-time balancing mechanisms exist in other systems under different terminology and market arrangements.}
These variables are directly related to the balancing-energy price component $P_t^{\mathrm{bal}}$. 

The second group contains market-reference features, including day-ahead prices, intraday 15-min and 60-min price indices, and the corresponding liquidity measures. 

These variables construct the market-reference component $P_t^{\mathrm{mkt}}$.

The third group contains system imbalance $V_t$, which falls under the scarcity-function feature category. 
This variable determines both the direction and magnitude of the scarcity component $P_t^{\mathrm{scar}}$. 
\subsection{Empirical Data Analysis}

\textbf{Temporal evolution of imbalance features: }
Fig.~\ref{fig:data_analysis} \textbf{(a)} shows that imbalance prices exhibit pronounced long-tailed behavior, especially during the energy crisis period in 2022, before becoming more stable in later years. 
In contrast, Fig.~\ref{fig:data_analysis} \textbf{(b)} shows that system imbalance remains more concentrated around zero, although its distribution also changes across years, with a slightly more negative imbalance pattern in 2025, reflecting increased renewable integration.

\textbf{Correlation between imbalance features: }
Fig.~\ref{fig:data_analysis} \textbf{(c)} and~\textbf{(d)} report Pearson and Spearman correlation magnitudes among $P_t$, the three price components, and $V_t$. 
The Pearson correlations in Fig.~\ref{fig:data_analysis} \textbf{(c)} show that $P_t$ is most strongly linearly associated with $P_t^{\mathrm{bal}}$, while its linear dependence on $P_t^{\mathrm{mkt}}$, $P_t^{\mathrm{scar}}$, and $V_t$ is weaker. 
In contrast, the Spearman correlations in Fig.~\ref{fig:data_analysis} \textbf{(d)} are stronger across most variable pairs, indicating nonlinear dependence induced by the imbalance pricing rules.

\textbf{Seasonal delivery-time patterns: }
Fig.~\ref{fig:data_analysis} \textbf{(e)}--\textbf{(h)} show average imbalance prices and system imbalance across delivery times for spring, summer, autumn, and winter, respectively, in 2025. 
The seasonal profiles reveal clear patterns: spring and summer show similar ranges, with summer exhibiting strong negative price movements around noon, while autumn and winter are more associated with positive imbalance-price levels, especially during afternoon and evening periods. 
These seasonal patterns motivate an evaluation protocol whose testing periods collectively cover a full year, ensuring that model performance is assessed across different seasonal conditions.

\textbf{Volatility of imbalance prices: }
Fig.~\ref{fig:data_analysis} \textbf{(i)} compares day-ahead, intraday, and imbalance prices averaged across 2022--2025. 
The observed volatility increases from day-ahead to intraday prices and is highest for imbalance prices. 
In particular, imbalance prices exhibit frequent extreme spikes in both directions, reflecting stronger volatility and tail risk. 
This highlights the difficulty of imbalance price forecasting and motivates the use of probabilistic models that can capture extreme price movements.

\section{Imbalance Pricing Rules}
\label{sec:prelim}
\textcolor{black}{This section introduces the imbalance pricing rules that map the raw feature vector $F_t$ into the imbalance price $P_t$. These rules provide the structural prior that will be embedded into the proposed MRINN model in Section~\ref{MRINN}.}

We consider quarter-hourly settlement periods indexed by $t$ and denote by $V_t$ the system imbalance (in MWh). The imbalance price $P_t$ is defined as the extremum over three price components, with the choice of $\min(\cdot)$ or $\max(\cdot)$ determined by the sign of $V_t$:
{\small
\begin{equation}
P_t \;=\;
\begin{cases}
\min\!\big(P^{\mathrm{bal}}_t,\; P^{\mathrm{mkt}}_t,\; P^{\mathrm{scar}}_t\big), & V_t < 0,\\[2pt]
\max\!\big(P^{\mathrm{bal}}_t,\; P^{\mathrm{mkt}}_t,\; P^{\mathrm{scar}}_t\big), & V_t \ge 0,
\end{cases}
\label{eq:imbalance_price_extremum}
\end{equation}
}
where $P^{\mathrm{bal}}_t$ is the balancing-energy-based price component (Eq.~\ref{eq:pbal}),
$P^{\mathrm{mkt}}_t$ is an exchange-based market reference price component (Eq.~\ref{eq:pmkt_final}),
and $P^{\mathrm{scar}}_t$ is a scarcity-function price component (Eq.~\ref{eq:pscar_final}).
Each component is specified in the following subsections.

\subsection{Balancing-Energy Price Component}
This component links the imbalance price to the balancing-energy outcome in period $t$ and depends on the activation direction and the sign of the system imbalance $V_t$:
{\small
\begin{equation}
P^{\mathrm{bal}}_t \;=\;
\begin{cases}
P^{\mathrm{act},+}_t, & I^{+}_t=1,\; I^{-}_t=1,\; V_t\ge 0,\\
P^{\mathrm{act},-}_t, & I^{+}_t=1,\; I^{-}_t=1,\; V_t< 0,\\
P^{\mathrm{act},+}_t, & I^{+}_t=1,\; I^{-}_t=0,\\
P^{\mathrm{act},-}_t, & I^{+}_t=0,\; I^{-}_t=1,\\
P^{\mathrm{VoAA},+}_t, & I^{+}_t=0,\; I^{-}_t=0,\; V_t\ge 0,\\
P^{\mathrm{VoAA},-}_t, & I^{+}_t=0,\; I^{-}_t=0,\; V_t< 0.
\end{cases}
\label{eq:pbal}
\end{equation}
}
where $P^{\mathrm{act},+}_t$ and $P^{\mathrm{act},-}_t$ denote the activation-weighted balancing-energy prices, as defined in Eq. \eqref{eq:pact_pos}--\eqref{eq:pact_neg}. $P^{\mathrm{VoAA},+}_t$ and $P^{\mathrm{VoAA},-}_t$ denote the values of avoided activation (VoAA) and are computed as the lowest and highest offer prices in the local positive and local negative aFRR balancing-energy merit-order lists for quarter-hour $t$, respectively.
$I^{+}_t$ and $I^{-}_t$ indicate whether positive or negative activation occurs, as defined in Eq. \eqref{eq:ind_pos}--\eqref{eq:ind_neg}.

{\small
\begin{equation}
P^{\mathrm{act},+}_t
= \frac{E^{\mathrm{aFRR},+}_t P^{\mathrm{aFRR},+}_t + E^{\mathrm{mFRR},+}_t P^{\mathrm{mFRR},+}_t}
{E^{\mathrm{aFRR},+}_t + E^{\mathrm{mFRR},+}_t}.
\label{eq:pact_pos}
\end{equation}
}

{\small
\begin{equation}
P^{\mathrm{act},-}_t
= \frac{E^{\mathrm{aFRR},-}_t P^{\mathrm{aFRR},-}_t + E^{\mathrm{mFRR},-}_t P^{\mathrm{mFRR},-}_t}
{E^{\mathrm{aFRR},-}_t + E^{\mathrm{mFRR},-}_t}.
\label{eq:pact_neg}
\end{equation}
}

{\small
\begin{equation}
I^{+}_t = \mathbb{I}\!\left[E^{\mathrm{aFRR},+}_t + E^{\mathrm{mFRR},+}_t > 0\right].
\label{eq:ind_pos}
\end{equation}
}

{\small
\begin{equation}
I^{-}_t = \mathbb{I}\!\left[E^{\mathrm{aFRR},-}_t + E^{\mathrm{mFRR},-}_t > 0\right].
\label{eq:ind_neg}
\end{equation}
}
where $E^{\mathrm{aFRR},\pm}_t$ and $E^{\mathrm{mFRR},\pm}_t$ are the activated aFRR and mFRR balancing-energy volumes (MWh), respectively. $P^{\mathrm{aFRR},\pm}_t$ and $P^{\mathrm{mFRR},\pm}_t$ are the corresponding 
prices (\euro/MWh). 
$\mathbb{I}[\cdot]$ is the indicator function.

\subsection{Market-Reference Price Component}
This component constructs an exchange-based reference for quarter-hour $t$ by forming a liquidity-dependent weighted sum of ramp-adjusted intraday and day-ahead indices:
{\small
\begin{equation}
P^{\mathrm{mkt}}_t
= w^{\mathrm{ID15}}_t\, P^{\mathrm{ID15,mkt}}_t
 + w^{\mathrm{ID60}}_t\, P^{\mathrm{ID60,mkt}}_t
 + w^{\mathrm{DA}}_t\, P^{\mathrm{DA,mkt}}_t .
\label{eq:pmkt_final}
\end{equation}
}
where $P^{\mathrm{ID15,mkt}}_t$, $P^{\mathrm{ID60,mkt}}_t$, and $P^{\mathrm{DA,mkt}}_t$ are defined in \eqref{eq:pid15_marked}--\eqref{eq:pda_marked}, and the weighting factors $w^{\mathrm{ID15}}_t$, $w^{\mathrm{ID60}}_t$, and $w^{\mathrm{DA}}_t$ are defined in \eqref{eq:w_id15}--\eqref{eq:w_da}.

{\small\begin{equation}
P^{\mathrm{ID15,mkt}}_t
= P^{\mathrm{ID15}}_t + \mathrm{ramp}(V_t)\cdot \max\!\Big(\text{C}_1,\ \text{C}_0\,|P^{\mathrm{ID15}}_t|\Big).
\label{eq:pid15_marked}
\end{equation}
}

{\small
\begin{equation}
P^{\mathrm{ID60,mkt}}_t
= P^{\mathrm{ID60}}_t + \mathrm{ramp}(V_t)\cdot \max\!\Big(\text{C}_2,\ \text{C}_0\,|P^{\mathrm{ID60}}_t|\Big).
\label{eq:pid60_marked}
\end{equation}
}

{\small
\begin{equation}
P^{\mathrm{DA,mkt}}_t
= P^{\mathrm{DA}}_t + \mathrm{ramp}(V_t)\cdot \max\!\Big(\text{C}_3,\ \text{C}_0\,|P^{\mathrm{DA}}_t|\Big).
\label{eq:pda_marked}
\end{equation}
}
where $P^{\mathrm{ID15}}_t$ is the intraday price index for 15-min products,
$P^{\mathrm{ID60}}_t$ is the corresponding intraday price index for 60-min products, and $P^{\mathrm{DA}}_t$ is the day-ahead market coupling price (all in \euro/MWh). The constants $\text{C}_0 = 0.1$, $\text{C}_1= 5$ (\euro/MWh), $\text{C}_2= 10$ (\euro/MWh), and $\text{C}_3= 15$ (\euro/MWh) are predefined parameters, and 
the $\mathrm{ramp}$ function is defined as:
{\small
\begin{equation}
\mathrm{ramp}(V_t) =
\begin{cases}
-1, & V_t < -\text{C}_4,\\
\dfrac{V_t}{\text{C}_4}, & -\text{C}_{4} \le V_t \le \text{C}_{4},\\
+1, & V_t > \text{C}_{4}.
\end{cases}
\label{eq:ramp}
\end{equation}
}
where $\text{C}_4 = 50 \text{ (MW)}$ is another constant and controls the width of the linear transition around $V_t\approx 0$.

{\small
\begin{equation}
w^{\mathrm{ID15}}_t = \min\!\left(1,\frac{L^{\mathrm{ID15}}_t}{\text{C}_5}\right),
\label{eq:w_id15}
\end{equation}
}

{\small
\begin{equation}
w^{\mathrm{ID60}}_t = \min\!\left(1-w^{\mathrm{ID15}}_t,\frac{L^{\mathrm{ID60}}_t}{\text{C}_6}\right),
\label{eq:w_id60}
\end{equation}
}

{\small
\begin{equation}
w^{\mathrm{DA}}_t = 1 - w^{\mathrm{ID15}}_t - w^{\mathrm{ID60}}_t,
\label{eq:w_da}
\end{equation}
}
where $L^{\mathrm{ID15}}_t$ and $L^{\mathrm{ID60}}_t$ denote the traded volumes used to assess intraday liquidity, and $\text{C}_5 = 200 \text{ (MW)}$ and $\text{C}_6 = 200 \text{ (MW)}$ are predefined constants for liquidity indication.

\subsection{Scarcity-Function Price Component}
This component introduces a scarcity adjustment as a signed cubic function of the absolute system imbalance $|V_t|$:
{\small
\begin{equation}
P^{\mathrm{scar}}_t =
\begin{cases}
P^{\mathrm{base}}_t, & |V_t|\le \text{C}_7,\\[4pt]
P^{\mathrm{base}}_t
+ \mathrm{sgn}(V_t)\,\text{C}_{10}
\left(\dfrac{|V_t|-\text{C}_7}{\text{C}_9-\text{C}_7}\right)^{3},
& \text{C}_7 < |V_t| \le \text{C}_8,\\[10pt]
P^{\mathrm{base}}_t
+ \mathrm{sgn}(V_t)\,\text{C}_{10}
\left(\dfrac{\text{C}_8-\text{C}_7}{\text{C}_9-\text{C}_7}\right)^{3},
& |V_t| > \text{C}_8.
\end{cases}
\label{eq:pscar_final}
\end{equation}
}
where $\mathrm{sgn}(V_t)\in\{-1,+1\}$ denotes the sign of the system imbalance. $\text{C}_7=200 \text{ (MW)}$  is the deadband threshold (no scarcity adjustment for $|V_t|\le \text{C}_7$), $\text{C}_8 = 800 \text{ (MW)}$  is the cap threshold beyond which the scarcity adjustment saturates, $\text{C}_9= 1000 \text{ (MW)}$  is the imbalance intersection point shaping the cubic curve, and $\text{C}_{10} = 1000$ (\euro/MWh) is the price intersection point scaling the magnitude of the scarcity adjustment. $P^{\mathrm{base}}_t$ is an \emph{unramped} exchange reference price, defined similarly to \eqref{eq:pmkt_final} as:
{\begin{equation}
P^{\mathrm{base}}_t
= w^{\mathrm{ID15}}_t\, P^{\mathrm{ID15}}_t
+ w^{\mathrm{ID60}}_t\, P^{\mathrm{ID60}}_t
+ w^{\mathrm{DA}}_t\, P^{\mathrm{DA}}_t .
\label{eq:pmkt_unramped}
\end{equation}
}

\begin{figure*}[!t]
    \centering
    \includegraphics[width=1\textwidth]{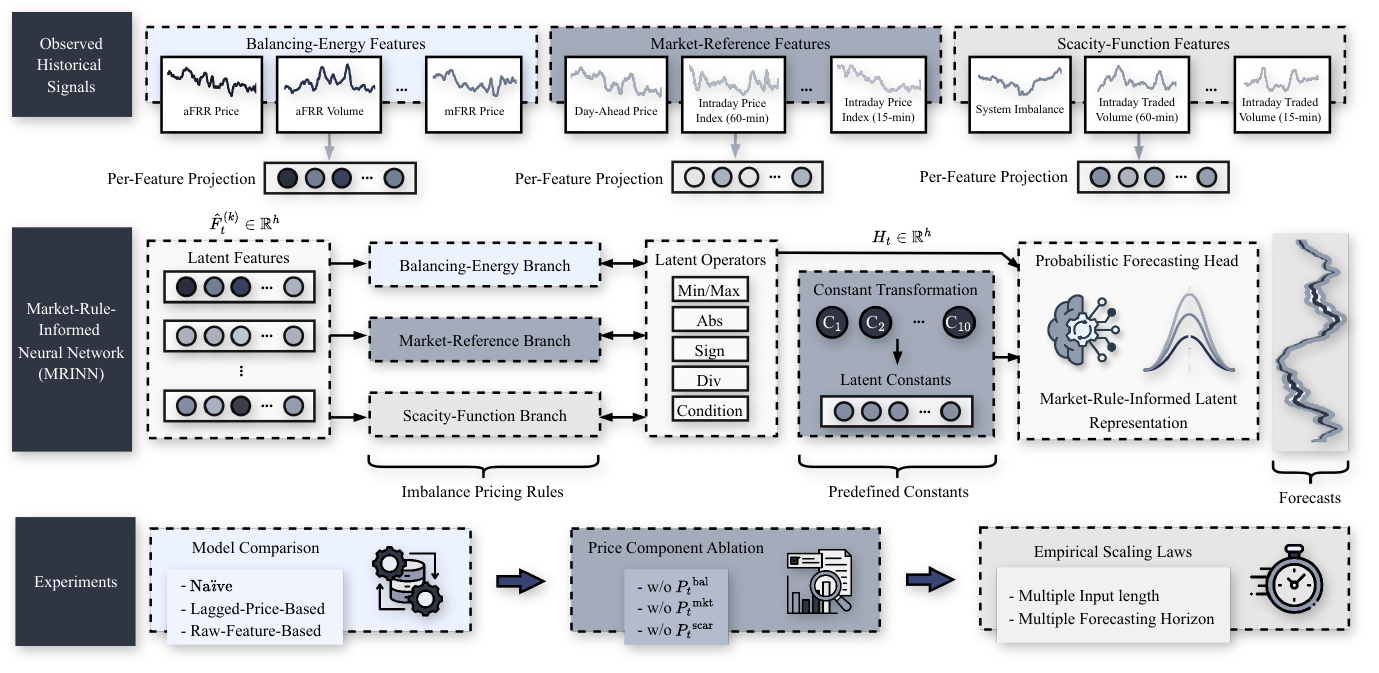}
    \caption{Structure illustration of the proposed MRINN and the corresponding experimental analysis.}
\label{fig:MRINN}
\end{figure*}

\section{Market-Rule-Informed Neural Network}
\label{MRINN}

\textcolor{black}{Since the imbalance price is generated by a known but non-invertible pricing map described from the last Section~\ref{sec:prelim}, directly compressing raw signals into the lagged price may discard predictive information. To preserve the original signal content while not losing the prior knowledge of pricing mechanisms,
we propose MRINN. }

Let $F_t^{(k)} \in \mathbb{R}$ denote the value of the $k$-th input feature at time step $t$, where $k \in \{1,\dots,K\}$ indexes the feature type and $K$ is the total number of input features. For example, $F_t^{(1)}$ may represent the system imbalance $V_t$. In the pricing formulation, the raw inputs at time $t$, namely $F_t^{(1)}, F_t^{(2)}, \dots, F_t^{(K)}$, are mapped to a scalar imbalance price $P_t \in \mathbb{R}$.

In detail, we first project each raw feature into a latent space of dimension $h$:

\begin{equation}
F_t^{(k)} \in \mathbb{R}
\xrightarrow{\text{Project}}
\hat{{F}}_t^{(k)} \in \mathbb{R}^h.
\end{equation}

After projection, each feature is represented as a latent vector, namely $\hat{F}_t^{(1)}, \hat{F}_t^{(2)}, \dots, \hat{F}_t^{(K)}$. Accordingly, the pricing mechanism is extended from \textbf{scalar operations} to \textbf{vector operations}.
In the latent space, these feature representations are passed through differentiable pricing-rule operators to obtain a latent representation of the imbalance price $H_t \in \mathbb{R}^h$, as illustrated in Fig.~\ref{fig:MRINN}.
However, when the pricing mechanism is implemented in the latent space, several of its original operations become challenging for backpropagation, as detailed in the following subsections (A--F).

\subsection{Min / Max Block}

The $\min$ and $\max$ operators are non-smooth, which may obstruct gradient flow during backpropagation in MRINNs. To address this, we replace the hard extremum operators with differentiable approximations. For two latent vectors $a,b \in \mathbb{R}^h$, note that the max operator can be rewritten as
\begin{equation}
\max(a,b) = b + \max(a-b, 0).
\end{equation}
We therefore approximate the hard $\max(\cdot,0)$ operation with the softplus function, where
\begin{equation}
\operatorname{softplus}(z)=\log(1+e^z).
\end{equation}
This gives the latent-space $\max$ block:
\begin{equation}
\operatorname{max}_{\mathrm{latent}}(a,b)
= b + \operatorname{softplus}(a-b).
\end{equation}
Similarly, the latent-space $\min$ block is defined by symmetry as
\begin{equation}
\operatorname{min}_{\mathrm{latent}}(a,b)
= -\operatorname{max}_{\mathrm{latent}}(-a,-b).
\end{equation}
All operations are applied element-wise. This yields differentiable and smooth approximations of the extremum operators appearing in Eq.~\ref{eq:imbalance_price_extremum}, Eq.~\ref{eq:pid15_marked}, Eq.~\ref{eq:pid60_marked}, Eq.~\ref{eq:pda_marked}, Eq.~\ref{eq:w_id15}, and Eq.~\ref{eq:w_id60}.

\subsection{Absolute Function Block}

As the absolute function $\lvert \cdot \rvert$ is non-differentiable at zero, we use the following differentiable approximation:
\begin{equation}
\operatorname{abs}_{\mathrm{latent}}(a)=\sqrt{a^2+\epsilon},
\end{equation}
where $\epsilon=10^{-7}$. This yields a differentiable approximation of $\operatorname{abs}$ used in Eq.~\ref{eq:pid15_marked}, Eq.~\ref{eq:pid60_marked}, Eq.~\ref{eq:pda_marked}, and Eq.~\ref{eq:pscar_final}.

\subsection{Sign Function Block}

As the $\operatorname{sign}(\cdot)$ operator is discontinuous at zero, we approximate it using the hyperbolic tangent function:
\begin{equation}
\operatorname{sign}_{\mathrm{latent}}(a)=\tanh(a),
\end{equation}
This function is fully differentiable and smoothly saturates toward $+1$ and $-1$, thereby mimicking the behavior of the original sign operator in Eq.~\ref{eq:pscar_final}.

\subsection{Division Block}

As division can become numerically unstable when the denominator approaches zero, we add a small positive constant to the denominator:
\begin{equation}
\operatorname{div}_{\mathrm{latent}}(a,b)=\frac{a}{b+\epsilon},
\end{equation}
This yields a numerically stable approximation of the division operation in Eq.~\ref{eq:pact_pos} and Eq.~\ref{eq:pact_neg} within the latent space.

\subsection{If-Else Condition Block}

Unlike decision trees or rule-based systems, standard neural networks do not naturally support hard if--else conditions. To address this, we replace discrete branching with a differentiable soft selection mechanism based on softmax weights.
Let us consider the following exemplary hard conditional rule:
\begin{equation}
\operatorname{cond}(A,B,C;a,b,c)=
\begin{cases}
A, & \text{if } a>b,\\
B, & \text{if } b>c,\\
C, & \text{otherwise},
\end{cases}
\end{equation}
where $a,b,c \in \mathbb{R}^h$ denote condition-related variables and $A,B,C \in \mathbb{R}^h$ denote the corresponding candidate outputs. To obtain a differentiable approximation, we first concatenate the condition variables and map them to branch-selection weights:
\begin{equation}
(w_1,w_2,w_3)=\operatorname{softmax}\bigl(\phi([a;b;c])\bigr),
\end{equation}
where $\phi(\cdot)$ denotes a learnable transformation, $w_1,w_2,w_3 \in [0,1]$, and
\begin{equation}
w_1+w_2+w_3=1.
\end{equation}
The hard if--else rule is then replaced by the following latent-space condition block:
\begin{equation}
\operatorname{cond}_{\mathrm{latent}}(A,B,C;a,b,c)
= A \odot w_1 + B \odot w_2 + C \odot w_3,
\end{equation}
where $\odot$ denotes element-wise multiplication.
Given different inputs $a,b,c$, the softmax function learns to adapt the branch probabilities accordingly. For example, if the first branch is strongly favored, the weights may approach $w_1 \approx 1$ and $w_2,w_3 \approx 0$, so that $\operatorname{cond}_{\mathrm{latent}}(A,B,C;a,b,c)$ effectively passes only $A$ to the output. In this way, the original hard conditional logic is approximated by a differentiable block in the latent space, as used in Eq.~\ref{eq:imbalance_price_extremum}, Eq.~\ref{eq:pbal}, Eq.~\ref{eq:ramp}, and Eq.~\ref{eq:pscar_final}.

\subsection{Constant Transformation Block}

As the pricing mechanism is implemented in the latent space, the predefined constants $C_1,\dots,C_{10}$ from Section~\ref{sec:prelim} cannot be directly used. First, they are scalar quantities, whereas the latent-space pricing mechanism operates on vectors in $\mathbb{R}^h$. Second, they are specified in the original physical scale, whereas the corresponding input features are scaled before entering the latent representation.

To ensure consistency, we group training features with the same physical unit, concatenate them along the sample dimension, and fit a \texttt{RobustScaler} for each unit group using the training data only. Each predefined constant is then transformed using the scaler associated with its corresponding unit. Finally, the transformed constant is broadcast to dimension $h$ so that it can be used in latent-space vector operations.

\subsection{Forecasting Head}

To obtain reliable probabilistic forecasts, we equip MRINN with the hierarchical
quantile head proposed in~\cite{yu2025orderfusion}, which is designed to mitigate
quantile crossing. The final
latent representation $H_t \in \mathbb{R}^h$ is fed into a dense layer to produce the median forecast:
\begin{equation}
    \hat{y}_{t,0.50} = f_{0.50}(H_t).
\end{equation}

The remaining quantiles are then constructed hierarchically through non-negative
increments. For example, for the quantile set
$\mathcal{Q}=\{0.10,0.25,0.45,0.50,0.55,0.75,0.90\}$, the lower quantiles are
generated outward from the median as
\begin{equation}
\begin{aligned}
    \hat{y}_{t,0.45} &= \hat{y}_{t,0.50} - \operatorname{softplus}(f_{0.45}(H_t)), \\
    \hat{y}_{t,0.25} &= \hat{y}_{t,0.45} - \operatorname{softplus}(f_{0.25}(H_t)), \\
    \hat{y}_{t,0.10} &= \hat{y}_{t,0.25} - \operatorname{softplus}(f_{0.10}(H_t)).
\end{aligned}
\end{equation}
Similarly, the upper quantiles are generated outward from the median as
\begin{equation}
\begin{aligned}
    \hat{y}_{t,0.55} &= \hat{y}_{t,0.50} + \operatorname{softplus}(f_{0.55}(H_t)), \\
    \hat{y}_{t,0.75} &= \hat{y}_{t,0.55} + \operatorname{softplus}(f_{0.75}(H_t)), \\
    \hat{y}_{t,0.90} &= \hat{y}_{t,0.75} + \operatorname{softplus}(f_{0.90}(H_t)).
\end{aligned}
\end{equation}
where $f_{\tau}(\cdot)$ denotes a quantile-specific dense layer. Since each
increment is constrained to be non-negative, the resulting forecasts satisfy the
desired ordering
$\hat{y}_{t,0.10} \leq \hat{y}_{t,0.25} \leq \hat{y}_{t,0.45} \leq
\hat{y}_{t,0.50} \leq \hat{y}_{t,0.55} \leq \hat{y}_{t,0.75} \leq
\hat{y}_{t,0.90}$, thereby preventing quantile crossing and improving the
reliability of probabilistic forecasts.

\subsection{Optimization Objective}
Following prior works~\cite{yu2025orderfusion, yu2026pricefmfoundationmodelprobabilistic}, we optimize the model using the Average Quantile Loss (AQL). 
Specifically, AQL measures the accuracy of multiple predicted quantiles:
\begin{equation}
    \text{AQL} 
    = 
    \frac{1}{N|\mathcal{Q}|} 
    \sum_{i=1}^N 
    \sum_{\tau \in \mathcal{Q}}  
    L_\tau(y_i, \hat{y}_{i,\tau}),
\end{equation}
where \( y_i \) and \( \hat{y}_{i,\tau} \) denote the true price and the predicted price quantile at level \(\tau\), respectively. The quantile loss \( L_\tau \) is defined as:
\begin{equation}
    L_\tau(y_i, \hat{y}_{i,\tau}) = 
    \begin{cases} 
      \tau \cdot (y_i - \hat{y}_{i,\tau}), 
      & \text{if } y_i \geq \hat{y}_{i,\tau}, \\
      (1 - \tau) \cdot (\hat{y}_{i,\tau} - y_i), 
      & \text{otherwise}.
    \end{cases}
\end{equation}

\section{Model Comparison}
\label{baselines}
We benchmark against three categories of baselines: (i) na\"ive baselines, (ii) lagged-price-based baselines, and (iii) raw-feature-based baselines. The na\"ive baselines serve as a reference to assess whether more advanced models produce meaningful improvements over simple heuristics. The comparison between lagged-price-based and raw-feature-based baselines is used to examine whether compressing raw signals into the lagged imbalance price leads to a loss of predictive information. The hyperparameter optimization is described in Appendix~\ref{appendix:hyperparams}. To ensure a fair comparison, testing metrics are reported only for the configuration of each model that achieves the best validation performance.

\subsection{Na\"ive Models}

We include three \textbf{Na\"ive Baselines}. The first adopts the latest imbalance price $P_t$ as the forecast for the next settlement period $P_{t+15\mathrm{min}}$, corresponding to a persistence baseline. 
In addition, prior studies have identified the 15-min intraday price index\footnote{In our setting, the index is defined as the volume-weighted average price over the $3$-hour interval before delivery. Specifically, the $15$-min and $60$-min indices are computed from the $15$-min and $60$-min products, respectively.}
as a strong persistence benchmark~\cite{iv9}. For consistency, we also include 60-min index as an additional na\"ive benchmark.
To obtain probabilistic forecasts, we first compute the residuals on the training data as the differences between the true prices and the corresponding na\"ive point forecasts. These residuals are then grouped by delivery time, and their empirical percentiles are estimated separately. Finally, the resulting delivery-specific residual percentiles are added back to the na\"ive point forecasts to construct probabilistic forecasts.

\subsection{Lagged-Price-Based Models}

When using only the lagged imbalance price $P_t$ as input, neither explicit temporal nor spatial inductive bias from advanced models is required. We therefore consider \textbf{Linear Quantile Regression (LQR)} for linear modeling, and \textbf{XGBoost (XGB)} and \textbf{Multi-Layer Perceptron (MLP)} for non-linear modeling, where prior work reports that XGB outperforms both persistence benchmarks and LSTM-based models~\cite{iv11}, while another study shows that MLP achieves lower testing loss than Gated Recurrent Unit (GRU), Temporal Fusion Transformer (TFT), and N-BEATS~\cite{iv10}.

\subsection{Raw-Feature-Based Models}

When using raw input signals, more advanced deep learning models can be considered. 
In the context of imbalance price forecasting, prior studies report that the \textbf{Attention-Based Bidirectional LSTM (AttnBiLSTM)} outperforms several competitive baselines, including Auto-Regressive Moving Average (ARMA), Gradient Boosting Regression Tree (GBRT), and Bahdanau-Based Sequence-to-Sequence (B-Seq2Seq)~\cite{iv12}. Similarly, another study also shows that this variant also outperforms Temporal Convolutional Network (TCN), LightGBM, and Transformer-based models~\cite{deng2024seasonality}.
In addition, we include several recent advanced deep learning models that have achieved strong performance across a wide range of domains, including transportation, weather, and electricity, namely \textbf{iTransformer}~\cite{itransformer}, \textbf{PatchTST}~\cite{PatchTST}, \textbf{TimesNet}~\cite{TimesNet}, and \textbf{TimeXer}~\cite{TimeXer}.

\subsection{Forecasting Settings}
\label{sec:forecasting_settings}

The forecasting process follows a 15-minute rolling window. To ensure strict temporal causality and avoid data leakage, only information available up to time $t$ is used to forecast the target from $t+15\mathrm{min}$. 
Moreover, we adopt a 3-fold evaluation protocol. 
The first fold uses data from 2022-01-01 to 2024-09-01 for training, from 2024-09-01 to 2025-01-01 for validation, and from 2025-01-01 to 2025-05-01 for testing. 
For the second and third folds, the training period is expanded by 4 months, while the validation and testing periods are shifted forward by 4 months, such that the full testing periods collectively cover a full year.
To ensure robustness, we compute the mean from 5 random seeds and 3 folds for each model.

\section{Operational Sensitivity Analysis}
\label{operational}
Beyond overall forecasting accuracy, practical deployment requires understanding how the model behaves when input information is limited or the forecasting task becomes more challenging. 
We therefore analyze operational sensitivity through two complementary studies: a price-component contribution analysis and an empirical scaling-law analysis over input length and forecasting horizon.

\subsection{Price Component Contribution}

As described in Section~\ref{sec:prelim}, the imbalance price in Eq.~(2) is determined by three price components: $P_t^{\mathrm{bal}}$, 
$P_t^{\mathrm{mkt}}$, and $P_t^{\mathrm{scar}}$.
Although these components jointly determine the final imbalance price, their relative contribution to forecasting is not necessarily equal. It is therefore
important to understand how each rule-derived price component, together with its
associated input features, contributes to the predictive representation learned by
the model. To this end, we conduct a component-removal analysis. For each experiment, we remove one price-component branch from the latent pricing mechanism and use the
remaining components to forecast the future price.

\textbf{(i) Without balancing-energy component.}
The aFRR- and mFRR-related input features are removed, and the pricing rules in
Eq.~(3)--Eq.~(7) are not used to construct the latent representation of
$P_t^{\mathrm{bal}}$. The model therefore relies only on the latent representations of
$P_t^{\mathrm{mkt}}$ and $P_t^{\mathrm{scar}}$ to forecast $P_{t+15\mathrm{min}}$.

\textbf{(ii) Without market-reference component.}
The market-related price features are removed, and the pricing rules in
Eq.~(8)--Eq.~(15) are not used to construct the latent representation of
$P_t^{\mathrm{mkt}}$. The model therefore relies only on the latent representations of
$P_t^{\mathrm{bal}}$ and $P_t^{\mathrm{scar}}$ to forecast $P_{t+15\mathrm{min}}$.

\textbf{(iii) Without scarcity-function component.}
The price features and the scarcity knowledge (Eq.~(16)--Eq.~(17)) are not used to construct the latent representation of
$P_t^{\mathrm{scar}}$. The model therefore relies only on the latent representations of
$P_t^{\mathrm{bal}}$ and $P_t^{\mathrm{mkt}}$ to forecast $P_{t+15\mathrm{min}}$.

\subsection{Empirical Scaling Laws}

We further investigate empirical scaling laws with respect to input length and
forecasting horizon. At each forecasting time $t$, the model receives historical
features from $t-N$ to $t$ and predicts the target price at $t+M$, where $N$ denotes
the look-back window and $M$ denotes the forecasting horizon. We evaluate multiple
look-back windows, $N \in \{0, 60, 180, 1440\}$ minutes, and forecasting horizons,
$M \in \{15, 30, 45, 60, 120, 180, 360, 540, 720, 1080, 1440\}$ minutes. For each $(N,M)$ configuration, we report the testing metrics to characterize how
forecasting performance scales with both the amount of historical information and
the prediction horizon. 

\textcolor{black}{It is worth noting that, in real-time operation, some input signals observed at time $t$ may not be immediately available for forecasting due to communication delays, publication lags, or data-fetching latency. Therefore, increasing the forecasting horizon $M$ can also be interpreted as evaluating the effect of \textbf{delayed
input availability}. For example, when $N=0$ and $M=180$, the model uses only the signals available at time $t$ to forecast the imbalance price at $t+180$ minutes;
equivalently, this reflects a setting in which the most recent available input for forecasting the price at time $t$ is delayed by 180 minutes. }

\begin{figure*}[!t]
    \centering
    \includegraphics[width=1\textwidth]{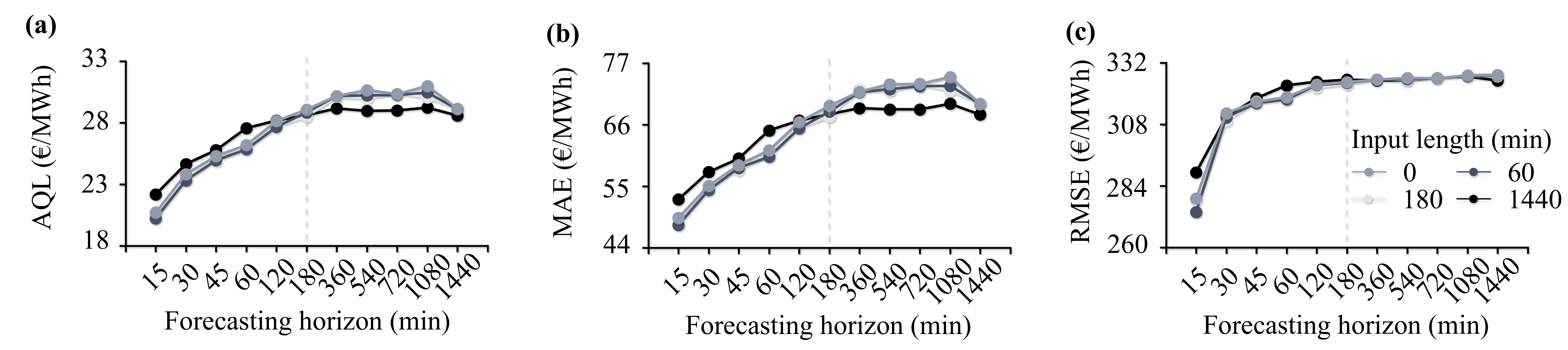}
    \caption{Experimental result analysis.
\textbf{(a)--(c)} Sensitivity analysis under different forecasting horizons and input lengths:
\textbf{(a)} AQL, \textbf{(b)} MAE, and \textbf{(c)} RMSE.}
    \label{fig:result}
\end{figure*}

\section{Experiment}
\label{experiment}
We conduct experiments to evaluate the proposed model from two perspectives. 
First, we compare it against na\"ive, lagged-price-based, and raw-feature-based baselines. 
Second, we perform an operational sensitivity analysis by varying input availability and characterizing empirical scaling laws with respect to input length and forecasting horizon.

\subsection{Model Comparison}

\textbf{Against Na\"ive Models.}
Table~\ref{tab:Forecasting} shows that the three Na\"ive baselines yield similarly poor probabilistic performance. In contrast, the proposed MRINN achieves an AQL of 20.70, which is \textbf{18.63\%} lower than the best Na\"ive baseline. A similar pattern is observed for point forecasting: MRINN reduces MAE from 56.29 to 49.36, corresponding to a \textbf{12.31\%} improvement, and reduces RMSE from 315.36 to 277.33, corresponding to a \textbf{12.06\%} improvement. These results confirm that simple persistence heuristics are insufficient for accurately modeling imbalance prices, especially under strong volatility and extreme price movements.

\textbf{Against Lagged-Price-Based Models.}
From Table~\ref{tab:Forecasting}, we observe that nonlinear models (XGB and MLP) clearly outperform the linear model (LQR), indicating that the relationship between historical prices and future imbalance prices is nonlinear. Meanwhile, the performance difference between XGB and MLP is small for both probabilistic and pointwise forecasting, suggesting that simply increasing nonlinear modeling capacity over lagged-price inputs provides limited additional gains. 
However, LQR has the lowest quantile-crossing violation among the lagged-price-based models, with an AQCR of only 0.62\%, and requires only 2 trainable parameters. Therefore, despite its lower forecasting accuracy, LQR can still be useful when quantile reliability is preferred and computational resources are extremely limited. 
Nevertheless, MRINN further improves over the strongest lagged-price-based baseline, reducing AQL from 22.51 to 20.70, corresponding to an \textbf{8.04\%} improvement. This further supports the non-invertibility issue discussed in the Introduction: aggregated lagged-price summaries lose information from raw features. Therefore, raw-feature-based baselines are needed to better capture the uncertainty of future imbalance price.

\textbf{Against Raw-Feature-Based Models.}
Table~\ref{tab:Forecasting} shows that all raw-feature-based models achieve similar predictive performance for both probabilistic forecasting (AQL) and pointwise forecasting (MAE and RMSE), with no statistically significant difference. However, these baselines still exhibit quantile crossing, with AQCR values ranging from 6.58\% to 8.53\%. In contrast, MRINN adopts the hierarchical quantile-head design, which guarantees zero quantile crossing. Moreover, the parameter comparison reveals a major advantage of MRINN: it requires only \textbf{1{,}817} trainable parameters\textcolor{black}{, and the resulting model size is only approximately \textbf{400 KB}, which is significantly smaller than typical neural-network forecasting models that can easily require hundreds of megabytes or even gigabytes when scaled to larger architectures. This efficiency advantage is particularly important for practical deployment, as energy trading increasingly involves battery assets and site-server deployment under resource-constrained operation, where memory, latency, and computational budgets can be limited. This advantage arises because MRINN explicitly incorporates imbalance pricing rules, allowing the model to represent the underlying pricing mechanism with much fewer parameters. In contrast, purely data-driven models must use {larger model capacity} to  \textbf{relearn} these rules, resulting in unnecessarily heavier models and higher computational cost.}

\begin{table}[t]
\centering
\caption{Forecasting Performance Comparison.}
\label{tab:Forecasting}
\begin{tabular}{
>{\raggedright\arraybackslash}p{1.55cm}
>{\centering\arraybackslash}p{0.9cm}
>{\centering\arraybackslash}p{1.1cm}
>{\centering\arraybackslash}p{0.9cm}
>{\centering\arraybackslash}p{1.1cm}
>{\centering\arraybackslash}p{0.9cm}
}
\toprule
\multirow{2}{*}{\textbf{Model}}
& \multicolumn{2}{c}{\cellcolor{citepurple!60}\textbf{Probabilistic}}
& \multicolumn{2}{c}{\cellcolor{citepurple!60}\textbf{Pointwise}}
& \multirow{2}{*}{\textbf{Params}} \\
\cmidrule(lr){2-3}
\cmidrule(lr){4-5}
& \textbf{AQL} 
& \textbf{AQCR}
& \textbf{MAE} 
& \textbf{RMSE} 
& \\

\midrule

Na\"ive$^{1}$      & 25.46 & \textcolor{gray}{0.00} & 56.29 & 321.84 & - \\
Na\"ive$^{2}$      & 25.49 & \textcolor{gray}{0.00} & 58.51 & 315.36 & - \\
Na\"ive$^{3}$      & 25.44 & \textcolor{gray}{0.00} & 58.53 & 316.77 & - \\

\midrule

LQR          & 24.48  & 0.62  & 53.30 & 294.26 &  2 \\
XGB          & 22.54  & 2.01 & 50.90  & 294.35 &  1.1k \\
MLP          & 22.51  & 1.89  & 50.87 & 294.02 &  8.9k \\

\midrule

PatchTST           & 20.72  & 6.77 & 49.43 & 277.73 & 17.9k  \\
iTransformer       & 20.71  & 8.44 & 49.38 & 277.32 &  13.6k \\
TimesNet           & 20.70  & 7.17 & 49.45 & 277.47 & 11.3k  \\
TimeXer            & 20.73  & 8.53 & 49.46 & 278.08 & 21.7k \\
AttnBiLSTM         & 20.72  & 6.58 & 49.34 & 277.51 & 28.3k \\

\midrule

\rowcolor{customgray!10}
\textbf{MRINN}   & \textbf{20.70} & \textbf{0.00} & \textbf{49.36} & \textbf{277.33} & \textbf{1.8k} \\

\bottomrule
\end{tabular}
\end{table}

\begin{table}[t]
\centering
\caption{Ablation Study of Price Component.}
\label{tab:price_component_ablation}
\begin{tabular}{
>{\raggedright\arraybackslash}p{1.55cm}
>{\centering\arraybackslash}p{0.9cm}
>{\centering\arraybackslash}p{1.1cm}
>{\centering\arraybackslash}p{0.9cm}
>{\centering\arraybackslash}p{1.1cm}
>{\centering\arraybackslash}p{0.9cm}
}
\toprule
\multirow{2}{*}{\textbf{Ablation}}
& \multicolumn{2}{c}{\cellcolor{citepurple!60}\textbf{Probabilistic}}
& \multicolumn{2}{c}{\cellcolor{citepurple!60}\textbf{Pointwise}}
& \multirow{2}{*}{\textbf{Rank}} \\
\cmidrule(lr){2-3}
\cmidrule(lr){4-5}
& \textbf{AQL} 
& \textbf{AQCR}
& \textbf{MAE} 
& \textbf{RMSE} 
& \\

\midrule
w/o $P_t^{\mathrm{bal}}$          & 22.63  & 0.00  & 54.22 & 288.33 & 2  \\
w/o $P_t^{\mathrm{mkt}}$         & 22.73  & 0.00 & 52.07 & 311.12 & 3  \\
w/o $P_t^{\mathrm{scar}}$         & 23.15  & 0.00  & 53.22 & 310.54 & 4  \\

\midrule
\rowcolor{customgray!10}
\textbf{All}   & \textbf{20.70} & \textbf{0.00} & \textbf{49.36} & \textbf{277.33} & \textbf{1} \\

\bottomrule
\end{tabular}
\end{table}

\subsection{Operational Sensitivity Analysis}

\textbf{Price Component Contribution.}
Table~\ref{tab:price_component_ablation} quantifies the contribution of each price component through a leave-one-component-out ablation study. Removing any component degrades both probabilistic and pointwise forecasting performance, confirming that $P_t^{\mathrm{bal}}$, $P_t^{\mathrm{mkt}}$, and $P_t^{\mathrm{scar}}$ provide complementary information. The largest AQL degradation occurs when removing $P_t^{\mathrm{scar}}$, with an \textbf{11.84\%} increase, indicating that the scarcity-function component contributes most to uncertainty modeling and tail-risk characterization. The largest MAE degradation occurs when removing $P_t^{\mathrm{bal}}$, with a \textbf{9.85\%} increase, suggesting that the balancing-energy component is most important for average pointwise accuracy. The largest RMSE degradation occurs when removing $P_t^{\mathrm{mkt}}$, with a \textbf{12.18\%} increase, suggesting that the market-reference component is especially important for controlling large pointwise errors and extreme deviations. Overall, the full model achieves the best performance, demonstrating the importance of jointly incorporating all three price components.

\textbf{Empirical Scaling Laws.}
Figs.~\ref{fig:result} \textbf{(a)-(c)} illustrate that two metrics degrade as the forecasting horizon increases, confirming that longer-horizon imbalance price forecasting is substantially more challenging. 
\textcolor{black}{This result is also relevant for industrial real-time deployment, since increasing the forecasting horizon can be interpreted as a proxy for delayed input availability: if the most recent signals are not available, the model must effectively forecast further ahead using older observations.}
\textbf{For short-term forecasting horizons} (below 180 min), shorter input windows (0 and 60 min), tend to achieve the best or near-best performance. In contrast, using a very long input length of 1440 min (1 day) leads to worse performance in this regime, suggesting that distant historical information may introduce noise that is less relevant for short-term prediction.
\textbf{For longer forecasting horizons} (beyond 180 min), the long input window (1440 min) becomes increasingly competitive. This trend suggests that longer-horizon prediction benefits more from low-frequency temporal structure, such as slower market dynamics and daily seasonality. In other words, while recent information dominates short-term prediction, long-range contextual information becomes more valuable as the forecasting target moves further into the future. This finding provides practical guidance for model deployment, as the input design can be adapted to the intended forecasting horizon to balance predictive accuracy and computational efficiency.

\section{Conclusion}
\label{conclusion}
This paper investigated probabilistic imbalance price forecasting under explicit market-rule-based price formation. Instead of treating imbalance prices as generic time-series targets, we examined whether known pricing rules should be integrated into neural forecasting models. The results show that compressing raw system and market signals into lagged imbalance prices can lead to information loss, while raw-feature-based models better preserve the information required for uncertainty-aware forecasting. More importantly, the proposed MRINN demonstrates that pricing rules should be used together with expressive neural networks: the rules provide structural inductive bias, while the neural network retains flexibility to model nonlinear  dependencies and predictive uncertainty.
\textcolor{black}{From an industrial deployment perspective, the proposed
framework supports resource-efficient and computationally
sustainable forecasting under operational constraints.}  
The price-component ablation further shows that balancing-energy, market-reference, and scarcity-function components provide complementary information, while the scaling-law analysis indicates that the optimal input length depends on the forecasting horizon. These results provide practical guidance for designing compact and deployable imbalance price forecasting models in operational energy systems.

A key limitation is that imbalance pricing rules are market-specific and may vary across bidding zones and settlement frameworks. As a result, the embedded rule structure must be adapted before transferring it to another market. Future work could investigate the reusability of differentiable operator blocks across markets and develop a foundation-model-style framework that integrates general data-driven representations with market-specific pricing mechanisms.

\section*{Appendix}

\subsection{Hardware and Computation}
\label{appendix:hardware}

We evaluate the computation of MRINN on both an \textbf{NVIDIA A100 GPU} and an \textbf{Intel Core i7-1265U CPU}. 
The training time is approximately 40 seconds on both hardware setups. 
Although GPUs generally provide faster numerical computation, the use of a Google Colab GPU introduces additional data-transfer overhead. 
Moreover, since MRINN is a lightweight model, the GPU is not fully utilized and therefore does not provide a significant acceleration over the CPU. 
The inference time is below \textbf{1 second} in both settings, indicating that MRINN is suitable for real-time applications.

\subsection{Hyperparameter Optimization}
\label{appendix:hyperparams}

For each model, we perform exhaustive hyperparameter optimization using grid search. 
The best model checkpoint is selected according to the validation performance and saved for final evaluation. 
The maximum number of training epochs is set to 70 for all neural models. 
We also evaluate batch sizes of 512, 1024, and 2048. 
Empirically, smaller batch sizes lead to earlier convergence in terms of epochs but require longer training time, whereas a batch size of 2048 does not consistently reach the best validation loss within 70 epochs. 
Therefore, we use a batch size of 1024 as a practical trade-off. 
For invalid hyperparameter combinations, the corresponding configurations are skipped during grid search. 
For example, in PatchTST, configurations where the patch length exceeds the input length, such as using $\texttt{patch\_len}=4$ when the input length is $2$, are excluded because they cannot form valid input patches. 
The full hyperparameter search spaces are summarized in Table~\ref{tab:hyperparams_all}.

\begin{table}[ht]
\begin{center}
\caption{Hyperparameter search space.}
\label{tab:hyperparams_all}
\begin{tabular}{ll}
\toprule
\textbf{Model} & \textbf{Search Space} \\
\midrule
MRINN &
\begin{tabular}[t]{@{}l@{}}
hidden\_size: \{8, 32, 128\} \\
n\_layers: \{2, 3, 4\} \\
\end{tabular}
\\
\midrule

iTransformer &
\begin{tabular}[t]{@{}l@{}}
hidden\_size: \{8, 32, 128\} \\
e\_layers: \{2, 3, 4\} \\
n\_heads: \{2, 4, 8\} \\
dropout: \{0.1, 0.3, 0.5\}
\end{tabular}
\\
\midrule

PatchTST &
\begin{tabular}[t]{@{}l@{}}
hidden\_size: \{8, 32, 128\} \\
e\_layers: \{2, 3, 4\} \\
n\_heads: \{2, 4, 8\} \\
dropout: \{0.1, 0.3, 0.5\} \\
patch\_len: \{2, 4, 8\}
\end{tabular}
\\
\midrule

TimesNet &
\begin{tabular}[t]{@{}l@{}}
hidden\_size: \{8, 32, 128\} \\
conv\_hidden\_size: \{8, 32, 128\} \\
e\_layers: \{2, 3, 4\} \\
dropout: \{0.1, 0.3, 0.5\}
\end{tabular}
\\
\midrule

TimeXer &
\begin{tabular}[t]{@{}l@{}}
hidden\_size: \{8, 32, 128\} \\
e\_layers: \{2, 3, 4\} \\
n\_heads: \{2, 4, 8\} \\
d\_ff: \{8, 32, 128\} \\
dropout: \{0.1, 0.3, 0.5\}
\end{tabular}
\\
\midrule

AttnBiLSTM &
\begin{tabular}[t]{@{}l@{}}
hidden\_size: \{8, 32, 128\} \\
layers: \{2, 3, 4\} \\
n\_heads: \{2, 4, 8\} 
\end{tabular}
\\

\bottomrule
\end{tabular}
\end{center}
\end{table}

\printbibliography[heading=bibintoc,title=Reference]

@article{bunn2021statistical,
  title={Statistical arbitrage and information flow in an electricity balancing market},
  author={Bunn, Derek W and Kermer, Stefan OE},
  journal={The Energy Journal},
  volume={42},
  number={5},
  pages={19--40},
  year={2021},
  publisher={SAGE Publications Sage CA: Los Angeles, CA}
}

@ARTICLE{8398478,
  author={Ng, Wing W. Y. and Zhang, Jianjun and Lai, Chun Sing and Pedrycz, Witold and Lai, Loi Lei and Wang, Xizhao},
  journal={IEEE Transactions on Industrial Informatics}, 
  title={Cost-Sensitive Weighting and Imbalance-Reversed Bagging for Streaming Imbalanced and Concept Drifting in Electricity Pricing Classification}, 
  year={2019},
  volume={15},
  number={3},
  pages={1588-1597},
  doi={10.1109/TII.2018.2850930}}

@ARTICLE{10418046,
  author={Shen, Xiaodong and Liu, Huixin and Qiu, Gao and Liu, Youbo and Liu, Junyong and Fan, Shixiong},
  journal={IEEE Transactions on Industrial Informatics}, 
  title={Interpretable Interval Prediction-Based Outlier-Adaptive Day-Ahead Electricity Price Forecasting Involving Cross-Market Features}, 
  year={2024},
  volume={20},
  number={5},
  pages={7124-7137},
  doi={10.1109/TII.2024.3355105}}

@ARTICLE{11456228,
  author={Yang, Changzhi and Pan, Huihui and Wang, Jue and Hong, Yuanduo},
  journal={IEEE Transactions on Industrial Informatics}, 
  title={DSFormer: Dual-Stream Transformers With Exogenous Variables for Electricity Price Forecasting}, 
  year={2026},
  volume={},
  number={},
  pages={1-10},
  doi={10.1109/TII.2026.3673412}}

@ARTICLE{8693845,
  author={Vu, Dao H. and Muttaqi, Kashem M. and Agalgaonkar, Ashish P. and Bouzerdoum, Abdesselam},
  journal={IEEE Transactions on Industrial Informatics}, 
  title={Short-Term Forecasting of Electricity Spot Prices Containing Random Spikes Using a Time-Varying Autoregressive Model Combined With Kernel Regression}, 
  year={2019},
  volume={15},
  number={9},
  pages={5378-5388},
  doi={10.1109/TII.2019.2911700}}

@inproceedings{iv2,
  title={Predictions of prices and volumes in the Nordic balancing markets for electricity},
  author={Backe, Stian and Riemer-S{\o}rensen, Signe and Bordvik, David A and Tiwari, Shweta and Andresen, Christian Andre},
  booktitle={2023 19th International Conference on the European Energy Market (EEM)},
  pages={1--6},
  year={2023},
  organization={IEEE}
}

@article{iv6,
  title={Interpretable probabilistic forecasting of imbalances in renewable-dominated electricity systems},
  author={Toubeau, Jean-Fran{\c{c}}ois and Bottieau, J{\'e}r{\'e}mie and Wang, Yi and Vall{\'e}e, Fran{\c{c}}ois},
  journal={IEEE Transactions on Sustainable Energy},
  volume={13},
  number={2},
  pages={1267--1277},
  year={2021},
  publisher={IEEE}
}

@article{iv9,
  title={Probabilistic forecasting of German electricity imbalance prices},
  author={Narajewski, Micha{\l}},
  journal={Energies},
  volume={15},
  number={14},
  pages={4976},
  year={2022},
  publisher={MDPI}
}

@article{iv10,
  title={Forecasting imbalance price densities with statistical methods and neural networks},
  author={Ganesh, Vighnesh Natarajan and Bunn, Derek},
  journal={IEEE Transactions on Energy Markets, Policy and Regulation},
  volume={2},
  number={1},
  pages={30--39},
  year={2023},
  publisher={IEEE}
}

@article{iv11,
  title={Prediction of Imbalance Prices through Gradient Boosting Algorithms: An Application to the Greek Balancing Market},
  author={Plakas, Konstantinos and Andriopoulos, Nikos and Papadaskalopoulos, Dimitrios and Birbas, Alexios and Housos, Efthymios and Moraitis, Ioannis},
  journal={IEEE Access},
  year={2025},
  publisher={IEEE}
}

@article{iv12,
  title={Interpretable transformer model for capturing regime switching effects of real-time electricity prices},
  author={Bottieau, Jeremie and Wang, Yi and De Greve, Zacharie and Vallee, Francois and Toubeau, Jean-Francois},
  journal={IEEE Transactions on Power Systems},
  volume={38},
  number={3},
  pages={2162--2176},
  year={2022},
  publisher={IEEE}
}

@article{deng2024seasonality,
  title={Seasonality in deep learning forecasts of electricity imbalance prices},
  author={Deng, Sinan and Inekwe, John and Smirnov, Vladimir and Wait, Andrew and Wang, Chao},
  journal={Energy Economics},
  volume={137},
  pages={107770},
  year={2024},
  publisher={Elsevier}
}

@article{o2024electricity,
  title={Electricity price forecasting in the irish balancing market},
  author={O’Connor, Ciaran and Collins, Joseph and Prestwich, Steven and Visentin, Andrea},
  journal={Energy Strategy Reviews},
  volume={54},
  pages={101436},
  year={2024},
  publisher={Elsevier}
}

@article{itransformer,
  title={iTransformer: Inverted Transformers Are Effective for Time Series Forecasting},
  author={Liu, Yong and Hu, Tengge and Zhang, Haoran and Wu, Haixu and Wang, Shiyu and Ma, Lintao and Long, Mingsheng},
  journal={arXiv preprint arXiv:2310.06625},
  year={2023}
}

@inproceedings{PatchTST,
  title     = {A Time Series is Worth 64 Words: Long-term Forecasting with Transformers},
  author    = {Nie, Yuqi and
               H. Nguyen, Nam and
               Sinthong, Phanwadee and 
               Kalagnanam, Jayant},
  booktitle = {International Conference on Learning Representations},
  year      = {2023}
}

@inproceedings{
TimesNet,
title={TimesNet: Temporal 2D-Variation Modeling for General Time Series Analysis},
author={Haixu Wu and Tengge Hu and Yong Liu and Hang Zhou and Jianmin Wang and Mingsheng Long},
booktitle={The Eleventh International Conference on Learning Representations },
year={2023},
url={https://openreview.net/forum?id=ju_Uqw384Oq}
}

@inproceedings{TimeXer,
 author = {Wang, Yuxuan and Wu, Haixu and Dong, Jiaxiang and Qin, Guo and Zhang, Haoran and Liu, Yong and Qiu, Yunzhong and Wang, Jianmin and Long, Mingsheng},
 booktitle = {Advances in Neural Information Processing Systems},
 editor = {A. Globerson and L. Mackey and D. Belgrave and A. Fan and U. Paquet and J. Tomczak and C. Zhang},
 pages = {469--498},
 publisher = {Curran Associates, Inc.},
 title = {TimeXer: Empowering Transformers for Time Series Forecasting with Exogenous Variables},
 url = {https://proceedings.neurips.cc/paper_files/paper/2024/file/0113ef4642264adc2e6924a3cbbdf532-Paper-Conference.pdf},
 volume = {37},
 year = {2024}
}

@misc{yu2026pricefmfoundationmodelprobabilistic,
      title={PriceFM: Foundation Model for Probabilistic Electricity Price Forecasting}, 
      author={Runyao Yu and Chenhui Gu and Jochen Stiasny and Qingsong Wen and Wasim Sarwar Dilov and Lianlian Qi and Jochen L. Cremer},
      year={2026},
      eprint={2508.04875},
      archivePrefix={arXiv},
      primaryClass={cs.CE},
      url={https://arxiv.org/abs/2508.04875}, 
}

@misc{yu2025orderfusion,
      title={OrderFusion: Encoding Orderbook for End-to-End Probabilistic Intraday Electricity Price Forecasting}, 
      author={Runyao Yu and Yuchen Tao and Fabian Leimgruber and Tara Esterl and Jochen Stiasny and Derek W. Bunn and Qingsong Wen and Hongye Guo and Jochen L. Cremer},
      year={2026},
      eprint={2502.06830},
      archivePrefix={arXiv},
      primaryClass={q-fin.CP},
      url={https://arxiv.org/abs/2502.06830}, 
}

@misc{runyaoEEM,
      title={Deep Learning for Electricity Price Forecasting: A Review of Day-Ahead, Intraday, and Balancing Electricity Markets}, 
      author={Runyao Yu and Derek W. Bunn and Julia Lin and Jochen Stiasny and Fabian Leimgruber and Tara Esterl and Yuchen Tao and Lianlian Qi and Yujie Chen and Wentao Wang and Jochen L. Cremer},
      year={2026},
      eprint={2602.10071},
      archivePrefix={arXiv},
      primaryClass={q-fin.CP},
      url={https://arxiv.org/abs/2602.10071}, 
}

@misc{runyaopscc,
      title={Orderbook Feature Learning and Asymmetric Generalization in Intraday Electricity Markets}, 
      author={Runyao Yu and Ruochen Wu and Yongsheng Han and Jochen L. Cremer},
      year={2026},
      eprint={2510.12685},
      archivePrefix={arXiv},
      primaryClass={q-fin.CP},
      url={https://arxiv.org/abs/2510.12685}, 
}

\end{document}